\documentclass[
 aip,
 apl,
 amsmath,amssymb,
 reprint,
]{revtex4-1}

\usepackage{graphicx}
\usepackage{dcolumn}
\usepackage{bm}

\begin{document}

\preprint{AIP/123-QED}

\title{Extensible 3D architecture for superconducting quantum computing}

\author{Qiang Liu}
\affiliation{National Laboratory of Solid State Microstructures, \\
School of Physics, Nanjing University, Nanjing 210093, China}
\author{Mengmeng Li}
\affiliation{National Laboratory of Solid State Microstructures, \\
School of Physics, Nanjing University, Nanjing 210093, China}
\author{Kunzhe Dai}
\affiliation{National Laboratory of Solid State Microstructures, \\
School of Physics, Nanjing University, Nanjing 210093, China}
\author{Ke Zhang}
\affiliation{National Laboratory of Solid State Microstructures, \\
School of Physics, Nanjing University, Nanjing 210093, China}
\author{Guangming Xue}
\affiliation{National Laboratory of Solid State Microstructures, \\
School of Physics, Nanjing University, Nanjing 210093, China}
\author{Xinsheng Tan}
\email{meisen0103@163.com}
\affiliation{National Laboratory of Solid State Microstructures, \\
School of Physics, Nanjing University, Nanjing 210093, China}
\author{Haifeng Yu}
\email{hfyu@nju.edu.cn}
\affiliation{National Laboratory of Solid State Microstructures, \\
School of Physics, Nanjing University, Nanjing 210093, China}
\author{Yang Yu}
\affiliation{National Laboratory of Solid State Microstructures, \\
School of Physics, Nanjing University, Nanjing 210093, China}

\date{\today}

\begin{abstract}
Using a multi-layered printed circuit board, we propose a 3D architecture suitable for packaging superconducting chips, especially chips that contain two-dimensional qubit arrays. In our proposed architecture, the center strips of the buried coplanar waveguides protrude from the surface of a dielectric layer as contacts. Since the contacts extend beyond the surface of the dielectric layer, chips can simply be flip-chip packaged with on-chip receptacles clinging to the contacts. Using this scheme, we packaged a multi-qubit chip and performed single-qubit and two-qubit quantum gate operations. The results indicate that this 3D architecture provides a promising scheme for scalable quantum computing.
\end{abstract}


\maketitle

Superconducting quantum circuits are promising candidates for realizing practical quantum computation\cite{You2003,You2005,Clarke2008,You2011,Buluta2011,Nation2012,Devoret2013,Xiang2013,Martinis2015}. Recently, quantum error correction with one-dimensional surface code protocol has been demonstrated on a nine-qubit superconducting quantum chip\cite{Barends2016,Kelly2016}. Given the current qubit control technique, thousands of qubits are needed to implement fault tolerant quantum computation\cite{Fowler2012,Ghosh2012,Barends2014,Riste2015,Billangeon2015,Versluis2016}. As the number of qubits increases and the microwave frequency pulses are commonly used for qubit control and measurement, conventional two-dimensional (2D) planar architectures are increasingly suffering from problems such as cross talks and cavity modes\cite{Wenner2011,Mutus2014,Averkin2014,Brecht2015}. There is therefore a lot of interest to extend the qubit control and measurement system to three-dimensional (3D) structures\cite{Bejanin2016,Brecht2016}. However, no quantum characteristic of the qubits has been reported so far. In this paper, we developed a convenient 3D architecture that transplants the control and measurement lines into a printed circuit board (PCB). Using this architecture, we do not need wire-bonding for packaging superconducting multi-qubit chips. We designed a thirteen-qubit superconducting chip and packaged it with our 3D architecture. Each qubit can be controlled and measured individually. The numerical simulation and experimental measurements show that electromagnetic properties of our scheme are comparable to those of wire-bonding. Moreover, we demonstrated coherent oscillation of single-qubit and two-qubit gate operations. Our results indicate that this packaging scheme is promising to support scalable quantum computation.

\begin{figure}[!htb]
\vspace{0.5cm}\includegraphics[width=8cm]{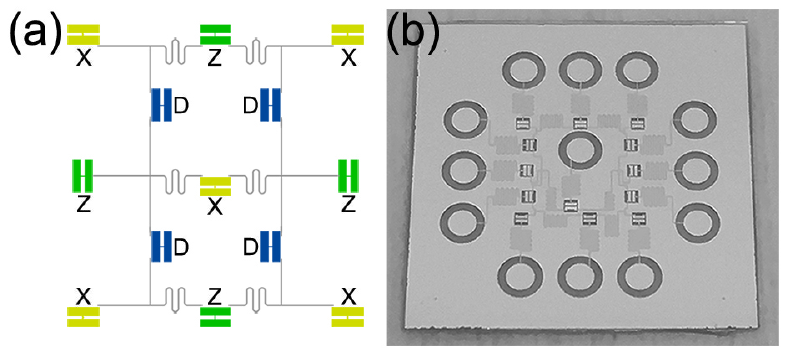}
\caption{(a) Schematic layout of a thirteen-qubit superconducting quantum chip compatible with surface code. Data qubits are labeled D, measure-X qubits are labeled X and measure-Z qubits are labeled Z. Gray lines represent couplers for coupling together nearest-neighbor qubits. (b) Photo of a 16$\times $16 mm$^{2}$ thirteen-qubit superconducting quantum chip. It contains thirteen transmon qubits coupled together with six CPW resonators. Each qubit has an individual CPW resonator for control and measurement, which also couples with circular pads (referred to as receptacles). }
\end{figure}

2D qubit arrays are often used to implement a surface code error correction algorithm\cite{Fowler2012,Chow2014,Kelly2015}. Each qubit couples to its nearest-neighbor qubits, forming a 2D ``square lattice". As shown in Fig.1(a), we designed a thirteen-qubit chip that is compatible with the 2D surface code protocol of quantum error correction. To avoid bias lines and simplify the system, we chose superconducting qubits as single Josephson junction transmons instead of frequency-tunable qubits. There are four data qubits (labeled D) and nine measurement qubits (X stands for measure-X qubits and Z stands for measure-Z qubits). To optimize coupler arrangement, we used six half-wavelength coplanar waveguide (CPW) resonators with branches (represented by gray lines) to couple together nearest-neighbor qubits\cite{Majer2007,Helmer2009}. Moreover, each qubit is designed to couple with an additional CPW resonator, which enables us to control and measure it with circuit quantum electrodynamics (cQED)\cite{Blais2004,Wallraff2004,You2003}. The size of the qubit chip is 16$\times $16 mm$^{2}$. As shown in Fig.1(b), thirteen circular pads located at the four sides and center of the chip are designed as receptacles for microwave contacts on the PCB. The qubit chip was fabricated on a 0.5 mm thick sapphire substrate. CPW resonators along with other big structures were patterned by ultraviolet lithography followed by a lift-off process. Every single Josephson junction (Al-AlO$_{x}$-Al) in parallel with planar capacitors was patterned by e-beam lithography and double evaporated with shadow evaporation.

\begin{figure}[!htb]
\vspace{0.5cm}\includegraphics[width=8cm]{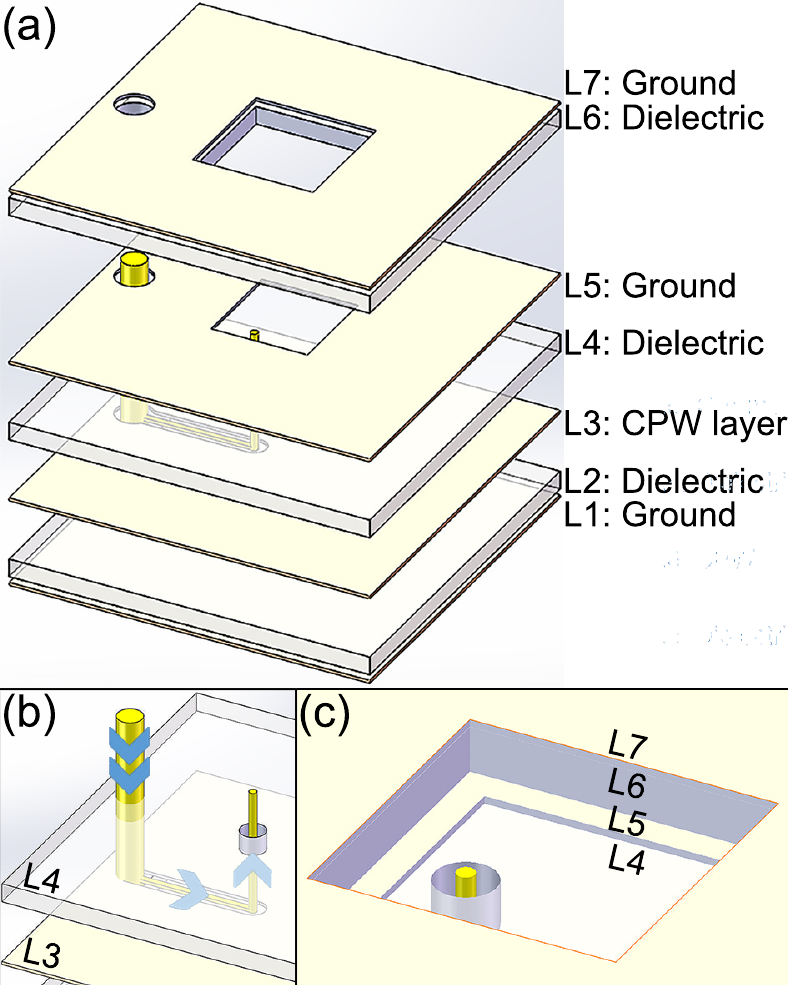}
\caption{(a) Schematic diagram of our 3D packaging architecture based on a PCB constituted of seven layers. Four metallic layers and three dielectric layers are labeled L1 to L7. (b) Zoomin of the key layers for microwave control lines. The thick column represents an inner conductor of the SMA connector, which would be soldered to the center strip of the CPW on L3. The latter protrudes from the surface of dielectric layer L4, acting as a contact. (c) The window for mounting the chip on L7 and L6 is 16.2$\times $16.2 mm$^{2}$, which is slightly bigger than the window on L5. The column threading L4 represents the contact. Since L6 plus L7 has almost the same thickness as our chip substrate, the superconducting quantum chip can be mounted face down in such a sink.}
\end{figure}

If conventional wire-bonding technique is used to package such a chip, we have to cross over bonding wires in order to access every wiring pads. Besides, long wires are needed to connect pads to conventional PCB ports, which may leak more electromagnetic energy and cause cross talks between different lines\cite{Johnson2010,Chow2011}. To solve these problems, we designed a 3D architecture for easily accessing pads at any location of the chip, especially those in the center. The 3D architecture is a multi-layered PCB. Shown in Fig.2 is its schematic diagram. For simplicity, we only draw one of thirteen microwave control lines. As depicted in Fig.2(a), the PCB contains seven layers labeled L1 to L7. There are four metallic layers alternated with three dielectric layers. The thickness of the four metallic layers are 35 $\mu$m (L1), 35 $\mu$m (L3), 87 $\mu$m (L5), and 35 $\mu$m (L7), respectively. All metallic layers are made of electrolytic copper with resistivity measured less than 1$\times$10$^{-7}$ $\Omega \cdot$m, which is close to that of pure copper. All dielectric layers are made of dielectric material RO4350B with a 0.508 mm thickness and a 3.66 relative dielectric constant. The thick column on the left represents the inner conductor of the SMA connector. It would be soldered together with the CPW center strip on L3. All CPWs are printed on this layer with impedance 50 $\Omega$, and they are covered by two adjacent dielectric layers (L2 and L4). Two ground layers (L1 and L5) outside the dielectric layers provide electrical isolation. The ground part of the CPW and all other ground layers are connected through many vias (not shown). The center strip of the CPW protrudes from the dielectric layer surface through via, forming contact to touch the receptacle on the superconducting quantum chip, as shown in Fig.2(b). Arrows represent the transmission path of input microwave signals. Under the protection of two ground layers and two dielectric layers, different CPWs are well isolated. Shown in Fig.2(c) is the window for mounting the chip on layers L7 and L6. The dimension of the window is 16.2$\times $16.2 mm$^{2}$, which is bigger than the window on layer L5. The thin column threading L4 represents the contact. Since the thickness of L6 plus L7 is almost the same as that of the chip substrate, the superconducting quantum chip can be tightly mounted face down in the sink. We obtain large design flexibility by either changing the location of the contacts or receptacles, which overcomes the difficulty of accessing pads at various locations of the superconducting quantum chips. In addition, this architecture is cheap and easy to manufacture. Therefore, it is a promising scheme to package multi-qubit superconducting quantum chips.

\begin{figure}[!htb]
\vspace{0.5cm}\includegraphics[width=8cm]{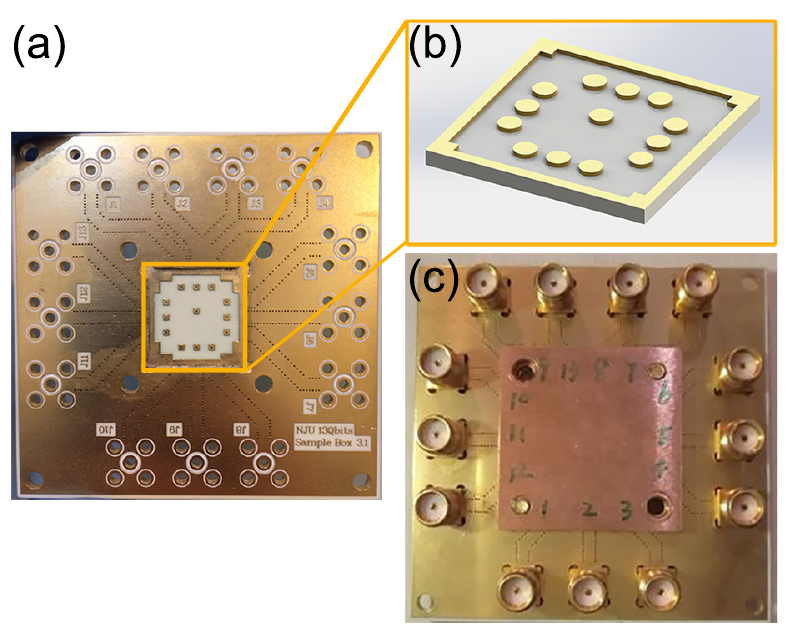}
\caption{(a) A photographic top-view of the PCB. L4, L5, and L7 along with thirteen contacts can be seen. (b) Schematic of the center region. Thirteen contacts are surrounded by ground edge (L5) with the same height above the white dielectric surface (L4). (c) Photo of packaged qubit chip assembly with SMA connectors soldered.}
\end{figure}

In Fig.3 we show a top-view of the PCB and the assembly of the whole package. The dimension of the PCB is 60$\times $60$\times $2 mm$^{3}$. The white area in Fig.3(a) is the exposed dielectric layer L4. The outmost ground layer L7 is gold plated. At each side of the PCB we leave spaces for mounting SMA connectors. There are thirteen SMA connectors for this chip. The center square region is 16.2$\times $16.2 mm$^{2}$ for accommodating and clamping the chip, which can be diced to fit the size of the sink. Fig.3(b) shows a schematic view of this region. There are thirteen contacts each covering a 1$\times$1 mm$^{2}$ area. They extend beyond the surface of the white dielectric layer (L4) by $\sim$0.1 mm. The surrounding ground edge (L5) has the same height as the contacts in order to form an even plane for the qubit chip. The qubit chip is put face down with receptacles touching the contacts. Indium foils can be put on the backside of the qubit chip, acting as thermal anchor and mechanical buffer. Then, two copper plates, which act as grounds, shields, and heat conductors, are clamped tightly with screws as shown in Fig.3(c).

We first checked the DC characterization of our PCB at room temperature. Using the standard four-point measurement method, we measured the resistance $R_s$ of the longest CPW strip in layer L3, which has a length $l$ = 28.9 mm and cross section area $S$ = 0.5$\times$0.035 mm$^2$. $R_s$ was measured to be 0.15 $\Omega$. The resistivity $\rho = R_s S/l$ is 9$\times$10$^{-8}$ $\Omega \cdot$m. Furthermore, we connected two CPW strips in layer L3 and obtained the contact resistance between PCB contacts and on-chip receptacles $R_c$ = 50 m$\Omega$. The small contact resistance indicates we can further shrink the size of the contact area.

\begin{figure}[!htb]
\vspace{0.5cm}\includegraphics[width=8cm]{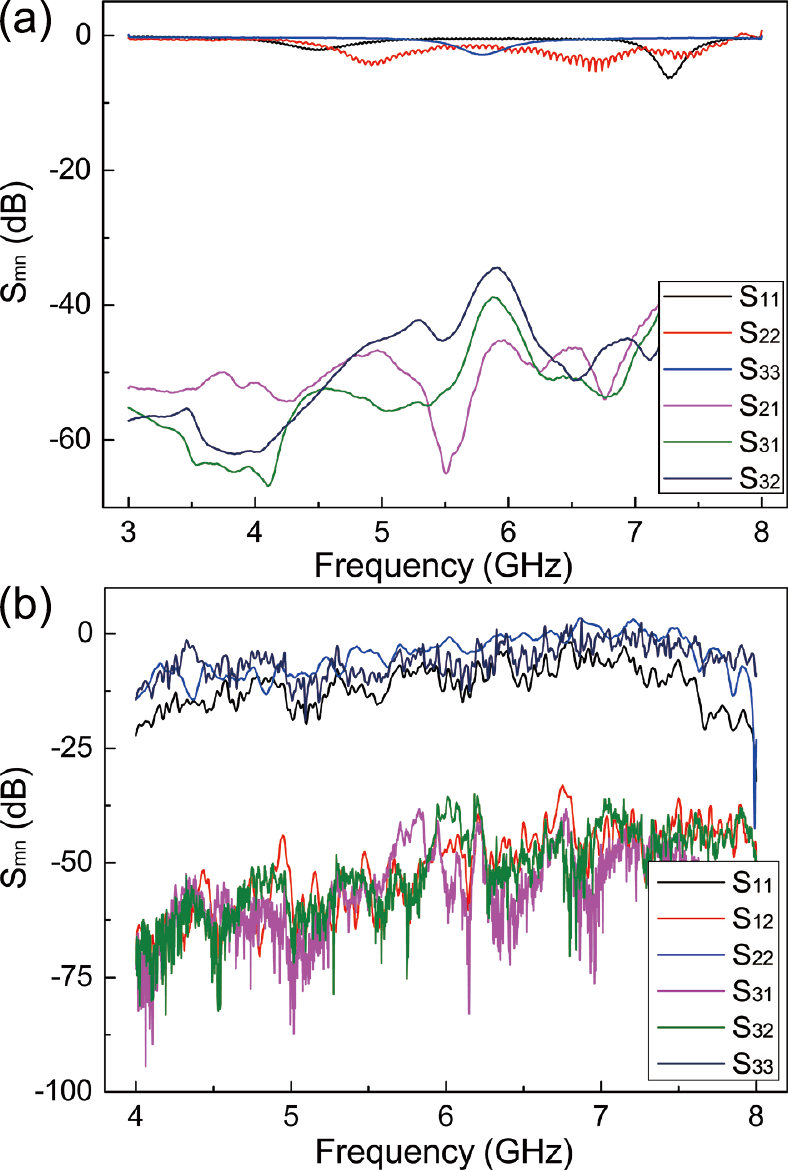}
\caption{Measured transmission curve among three different ports at both room temperature (a) and 20 mK (b).}
\end{figure}

It is well known that cavity modes are a challenge to frequency-dependent microwave measurements. The frequency of cavity modes depends on the geometric size of the inner open space of the sample cell\cite{Averkin2014}. The inner space of our package can be easily designed to be smaller than 3 cm, which results in the lowest mode frequency to be over 10 GHz. Generally, the frequencies of qubits, control/measurement resonators, and couplers are all in the range of 3-8 GHz. Therefore, the effect of cavity modes can be neglected in our package. Another issue we have to pay attention to is reflection and loss caused by sharply angled structures. We simulated the transmission ($S_{21}$) of both straight CPW and CPW ended with a vertical via. As shown in Fig.2, the height of this via is about 0.508 mm, which is much shorter than the concerned wavelength. The simulation results indicate that although the vertical via introduces reflection structures, the intensity of the loss peaks is about 1.25 dB and not significant.

To fully characterize this packaging scheme, we simulated and measured the electromagnetic property of our PCB, especially the cross talks between different ports. Theoretically, two ground layers (L1 and L5) will confine the magnetic flux in the CPW layer (L3), which may cause one line to have a higher density (or chance) of affecting other lines and increase the cross talk compared to a normal CPW. To verify this, we simulated the electromagnetic transmission for both scenarios. They exhibit comparable cross talk values, which range from -40 dB to -60 dB. Cross talks were also measured among ports 1, 2, and 3. Port 1 sends signal to the center of the chip. Port 2 and port 3 are neighboring ports that send signals to the edge of the chip. We find that the isolation among the three ports is at the level of -40 dB to -50 dB at both room temperature (Fig.4(a)) and 20 mK (Fig.4(b)).

One of the advantages of our architecture is that it solves wire-bonding problems. Therefore, it is important to compare the isolation of our architecture with that of the conventional wire-bonding scenario. We simulated the transmission and reflection of two ports in conventional wire-bonding scenario. We find that its cross talk ($\sim-30$ dB) is worse than that of our scheme.

\begin{figure}[!htb]
\vspace{0.5cm}\includegraphics[width=8cm]{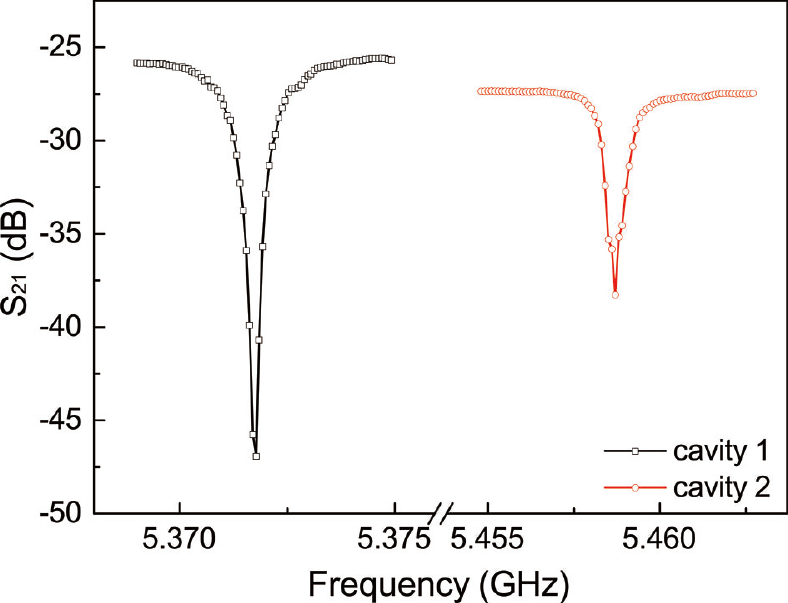}
\caption{Typical S$_{21}$ curves of two of thirteen CPW resonators. The resonators exhibit an internal quality factor of 62000 at 5.372 GHz and 13000 at 5.459 GHz. Devices were measured at 20 mK temperature.}
\end{figure}

To further characterize the performance of this packaging architecture, we cooled the qubit chip down to 20 mK in a dilution refrigerator (DR). The qubit chip was well protected from the external noise. We measured microwave transmission curves ($S_{21}$) by a network analyzer (Agilent E8363B). Fig.5 shows two typical transmission curves ($S_{21}$) of the CPW resonators on the qubit chip. Resonance dips with internal quality factor $Q_{i}$ of 62000 (cavity 1) and 13000 (cavity 2) at a high driven power (same power as that used to perform high-power-readout\cite{reed2010}) are at frequency 5.732 GHz (cavity 1) and 5.459 GHz (cavity 2). These values are comparable to those of planar wire-bonding chips.

\begin{figure}[!htb]
\vspace{0.5cm}\includegraphics[width=8cm]{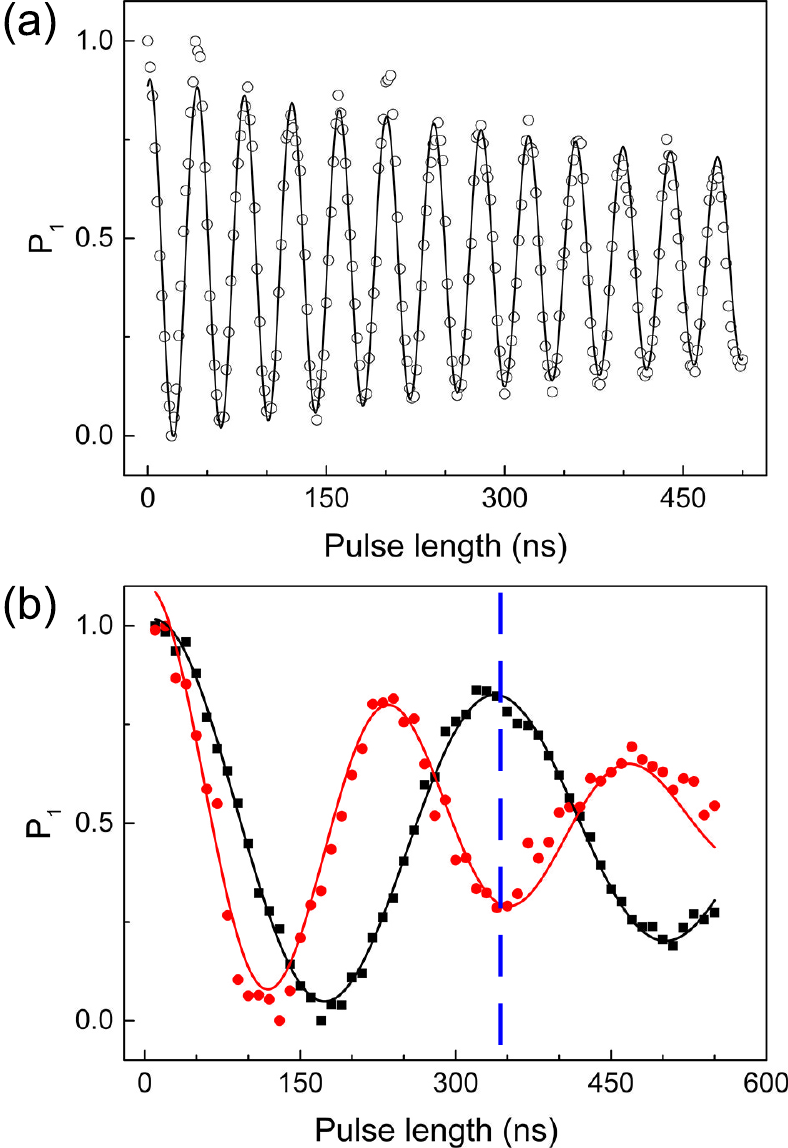}
\caption{(a) Typical Rabi oscillation of a single qubit. Open circles are data and solid line is the best fit. (b) Rabi oscillations of qubit-2 with different rates for the initial state of qubit-1 at $|0 \rangle$ (square) or at $|1 \rangle$ (circle), respectively. Solid dots are data and solid lines are the best fits. A CNOT gate is achieved when the cross-resonance pulse lasts for 350 ns (marked with dashed line).}
\end{figure}

To further benchmark this 3D architecture, we measured the quantum dynamics of our qubits. Using the cQED systems formed by the CPW resonators and transmon qubits, we demonstrated Rabi oscillations\cite{Koch2007}. The driving microwave pulses were sent to qubits through CPW resonators. After turning off the driving microwave, we measured the qubit states with a common microwave heterodyne setup\cite{Wallraff2004}. Shown in Fig.6(a) is a typical Rabi oscillation of a transmon qubit. The decoherence time is about 3.47 $\mu$s, which is similar to that of our previous transmon qubit chip packaged with conventional wire-bonding. We also demonstrated the two-qubit CNOT operation using the cross-resonance (CR) method\cite{Grajcar2006,Ashhab2006,Ashhab2007,Ashhab2008,Corcoles2013}. For two qubits coupled through a half-wavelength CPW resonator, CR arises when qubit-1 (control qubit) is driven at the frequency of qubit-2 (target qubit). To demonstrate the CNOT gate, we applied a shaped microwave $\pi $ pulse to the control qubit at the frequency of the target qubit. As shown in Fig.6(b), qubit-2 exhibits Rabi oscillations with different rates depending on the state of qubit-1. Square dots represent that qubit-1 is initially at $|0 \rangle$ state while circular dots represent qubit-1 is initially at $|1 \rangle$ state. We can see that if the CR pulse has a length about 350 ns, qubit-2 is at either $|0\rangle $ or $|1\rangle $ state depending on the initial state of qubit-1. Therefore, applying a CR pulse with 350 ns we can realize a CNOT gate. The estimated fidelity of the CNOT gate is about 0.63. In the single-qubit case, gate fidelity of the NOT gate is about 0.98, as calculated by the method of Horodecki and Nielsen\cite{Horodecki1999,Nielsen2002}. Currently, the performance of our 3D architecture is limited by the short decoherence time of our qubits (750 ns for squared dots, and 340 ns for circular dots in Fig.5(b)). But with the improvement of fabrication technique, this 3D architecture is a promising approach for scalable quantum computation.

Our architecture has the potential to scale to large numbers of qubits. In our scheme, each qubit has its own control and readout microwave line. For a moderate increase of qubits, we can avoid the cross of lines by optimizing the chip design. For instance, we can replace SMA connectors with compact SMP connectors. We may also reduce the size of PCB contacts and on-chip receptacles while increasing the size of the chip. Furthermore, we have wiring flexibility for our multi-layered PCB scheme. With an even larger number of qubits, we can add more CPW layers in the PCB. The in-between ground layers will minimize the cross talk of lines in different CPW layers.

In summary, we have proposed a 3D architecture based on multi-layered PCB technique. By using this scheme, we packaged two-dimensional qubit arrays without bonding wires. The measured quality factor of the CPW resonators is comparable to that of our previous wire-bonded qubit chips. The electromagnetic properties of our scheme are comparable to those of wire-bonding. Furthermore, one and two qubit operations with reasonable fidelity are demonstrated. Due to its flexibility for accessing on-chip pads at any location, this scheme is promising for salable quantum computation.

\textit{Supplementary materials} See Supplementary for more information about measurement system and numerical simulations.

\textit{Acknowledgments} This work was partly supported by the NKRDP of China (Grant No. 2016YFA0301802) and NSFC (Grant No. 11474152, No. 91321310, No.11274156, No.11504165, and No. 61521001).

\end{document}